\documentclass[reprint, floatfix, pra, nofootinbib, superscriptaddress]{revtex4-1}

\usepackage{graphicx}
\usepackage{bm}
\usepackage{physics}
\usepackage{multirow}
\usepackage[table,dvipsnames,svgnames]{xcolor}
\usepackage[linkcolor=cyan]{hyperref}
\usepackage{svg}
\usepackage{amsmath}
\usepackage{amsfonts}
\usepackage{float}
\usepackage{datetime}

\begin{document}

\title{Quantifying the limits of controllability for the nitrogen-vacancy electron spin defect}

\author{Paul Kairys}
\email{kairyspm@ornl.gov}
\affiliation{Mathematics and Computer Science Division, Argonne National Laboratory, Lemont, IL}
\affiliation{{Q-NEXT}, Argonne National Laboratory, Lemont, IL}

\author{Jonathan C. Marcks}
\affiliation{{Q-NEXT}, Argonne National Laboratory, Lemont, IL}
\affiliation{Material Science Division and Center for Molecular Engineering, Argonne National Laboratory, Lemont, IL}
\affiliation{Pritzker School of Molecular Engineering, University of Chicago, Chicago, IL}

\author{Nazar Delegan}
\affiliation{{Q-NEXT}, Argonne National Laboratory, Lemont, IL}
\affiliation{Material Science Division and Center for Molecular Engineering, Argonne National Laboratory, Lemont, IL}

\author{Jiefei Zhang}
\affiliation{{Q-NEXT}, Argonne National Laboratory, Lemont, IL}
\affiliation{Material Science Division and Center for Molecular Engineering, Argonne National Laboratory, Lemont, IL}

\author{David D. Awschalom}
\affiliation{{Q-NEXT}, Argonne National Laboratory, Lemont, IL}
\affiliation{Material Science Division and Center for Molecular Engineering, Argonne National Laboratory, Lemont, IL}
\affiliation{Pritzker School of Molecular Engineering, University of Chicago, Chicago, IL}
\affiliation{Department of Physics, University of Chicago, Chicago, IL}

\author{F. Joseph Heremans}
\affiliation{{Q-NEXT}, Argonne National Laboratory, Lemont, IL}
\affiliation{Material Science Division and Center for Molecular Engineering, Argonne National Laboratory, Lemont, IL}
\affiliation{Pritzker School of Molecular Engineering, University of Chicago, Chicago, IL}

\date{September, 2024}

\begin{abstract}
   Solid-state electron spin qubits, like the nitrogen-vacancy center in diamond, rely on control sequences of population inversion to enhance sensitivity and improve device coherence. But even for this paradigmatic system, the fundamental limits of population inversion and potential impacts on applications like quantum sensing have not been assessed quantitatively. Here, we perform high accuracy simulations beyond the rotating wave approximation, including explicit unitary simulation of neighboring nuclear spins. Using quantum optimal control, we identify analytical pulses for the control of a qubit subspace within the spin-1 ground state and quantify the relationship between pulse complexity, control duration, and fidelity. We find exponentially increasing amplitude and bandwidth requirements with reduced control duration and further quantify the emergence of non-Markovian effects for multipulse sequences using sub-nanosecond population inversion. From this, we determine that the reduced fidelity and non-Markovianity is due to coherent interactions of the electron spin with the nuclear spin environment. Ultimately, we identify a potentially realizable regime of nanosecond control duration for high-fidelity multipulse sequences. These results provide key insights into the fundamental limits of quantum information processing using electron spin defects in diamond.
\end{abstract}

\maketitle

\section{Introduction}

Defects in solid-state systems that host isolated electron spins are an extremely promising platform for quantum technologies \cite{wolfowicz_quantum_2021, awschalom_quantum_2018, pompili2021realization}. Arguably the most widely studied and technologically mature system is the nitrogen-vacancy (NV$^-$) center in diamond, owing to a number of convenient properties like room temperature coherent operation, magnetically driven dynamics, and optical readout. These properties have driven a number of applications for NV$^-$ center technologies in the areas of quantum information, including sensing and computing \cite{barry_sensitivity_2020,bradley_ten-qubit_2019,casola_probing_2018}. In all applications, maximizing coherence and operation fidelity is paramount.

The standard technique to extend qubit coherence relies on multipulse refocusing sequences to dynamically decouple the system from its environment \cite{viola_dynamical_1999,souza_robust_2011}. Defined for a two-level system, a multipulse sequence is composed of many population-inversions, commonly referred to as $\pi$-pulses because they generate a rotation of angle $\pi$ around an axis of the Bloch sphere. For applications in sensing or computing, it is important that these $\pi$-pulses are simultaneously fast, high-fidelity, and Markovian \cite{barry_sensitivity_2020,degen_quantum_2017}. If the $\pi$-pulses are instantaneous and have perfect fidelity this leads to a signal filtering formalism useful for many applications \cite{biercuk_dynamical_2011}. 

Unfortunately, in practical application, these requirements are contradictory because every non-trivial unitary evolution that can be generated in a quantum system has a finite control duration and finite fidelity loss \cite{loretz_spurious_2015,koch_quantum_2022}. Therefore it is crucial to understand the fundamental limits of population inversion in NV$^-$ center systems and the impact that these limits can have on application performance.

Assessing the limits of population inversion at short control duration can be formalized as a quantum optimal control (QOC) task. As pulse duration decreases, a larger drive field is necessary to generate the desired rotation, breaking common experimental control approximations like resonance conditions and the rotating wave approximation (RWA) \cite{fuchs_gigahertz_2009, zeuch_exact_2020}. Therefore, simulation accuracy is paramount to capture the relevant physics in this regime and accurately guide experimental effort.

\begin{figure*}[t]
    \centering
    \includegraphics[width=\textwidth]{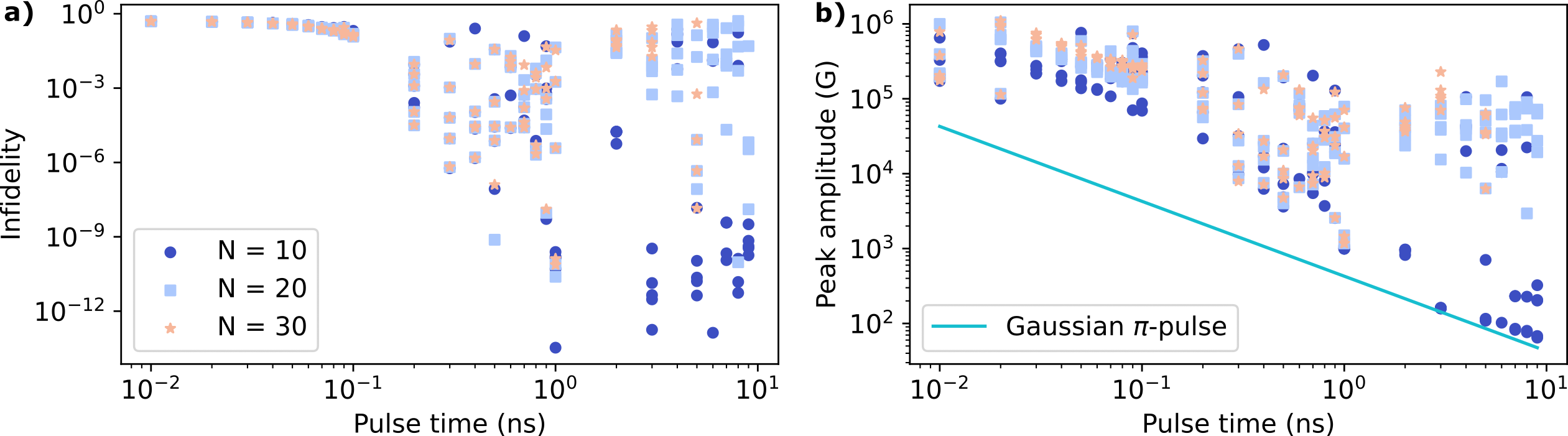}
    \caption{The numerical results for an ensemble of optimizations with a varying number of pulse basis functions, $N$, yielding a total number of $3N$ optimized parameters. \textbf{a)} shows the optimal final-time infidelity found for decreasing control duration. We observe an exponential increase in infidelity with decreasing control duration. In \textbf{b)} the maximum pulse amplitude is plotted as a function of decreasing control duration. For reference, the maximum amplitude required by a Gaussian envelope function with $\pi$ area for the same duration is plotted. This Gaussian reference is what is anticipated analytically in the long-time, low-amplitude regime where the RWA is valid and the NV$^-$ center electron system can be well described as a two-level system.}
    \label{fig:infidelity_amplitude}
\end{figure*}

In this work, we build an accurate model of a NV$^-$ center and neighboring nuclear spins system using parameters derived from detailed experiments \cite{bradley_ten-qubit_2019,abobeih_atomic-scale_2019}. We include significant non-secular terms in the spin Hamiltonian and model global driving conditions, while avoiding the RWA. This enables us to accurately and quantitatively probe the fundamental limits of information dynamics in the NV$^-$ center system.

We perform an ensemble of QOC simulations using our NV$^-$ center model. We examine how the fidelity of optimized $\pi$-pulses depends on pulse duration and identify the mechanism of control and the loss of fidelity at short pulse times. Then, we analyze the optimal pulses and quantify the growth in amplitude and bandwidth required to achieve population inversion. Using these results, we consider these pulses for a model multipulse quantum sensing application, enabling us to quantify the non-Markovian effects that emerge in multipulse sequences of short-time $\pi$-pulses. We conclude by discussing the feasibility of our identified controls and future research directions. 

The rest of this paper is structured as follows: in Section~\ref{sec:model} we define the model NV$^-$ center system and discuss the experimental context, thereafter we present our numerical results in Section~\ref{sec:numerical_results} and connect these results to an example quantum sensing application in Section~\ref{sec:application}. Finally, we state and discuss our conclusions in Section~\ref{sec:conclusion}. 

\section{System model}\label{sec:model}

The negatively charged NV$^-$ center in diamond hosts a number of coherent quantum degrees of freedom. The most commonly used is the ground state electronic spin-1 degree of freedom. This electron spin couples to nearby nuclear spins in diamond via the hyperfine interaction. The neighboring nuclei consist of the spin-1 nitrogen nucleus in the NV$^-$ center itself and randomly distributed spin-$1/2$ $^{13}$C nuclei. There is an additional diffuse bath of electron and nuclear spin-containing Nitrogen defects that we do not consider in this work. All of these spins couple to magnetic fields and can therefore be controlled by manipulating the external magnetic field. 

Ignoring interactions due to the electric field, the Hamiltonian of this NV$^-$ center system including a single $^{14}$N nuclear spin and multiple $^{13}$C spins is written as \cite{rembold_introduction_2020}:

\begin{align}
H(t) &= DS_z^2 + QI_{N,z}^2 + \sum_j \omega_j I_{C_j,z} \\
\nonumber &~~~~+  \sum_{j} \vec{S} \mathcal{N}_j \vec{I_j} + \vec{B}(t) \cdot \bigg[ \gamma_e \vec{S} + \sum_j \gamma_j \vec{I}_{j} \bigg],
\end{align}

where $\vec{S} = (S_x,S_y,S_z)$ is the vector of spin-1 operators that act on the electron spin of the NV$^-$ center and $\vec{I_j}=(I_{(j,x)},I_{(j,y)},I_{(j,z)})$ is the vector of spin operators acting on nuclei $j$. The parameters $D,Q,\omega_j$ are the zero-field splitting, nuclear quadrupole, and resonance frequencies of the electron, nitrogen nucleus, and carbon nuclei, respectively. The spin interactions are specified by the hyperfine tensor $\mathcal{N}$ and $\gamma_j$ are the nuclear gyromagnetic ratios that specify the interaction with the time-dependent magnetic field $\vec{B}(t)$. 

\begin{figure*}
    \centering
    \includegraphics[width=\textwidth]{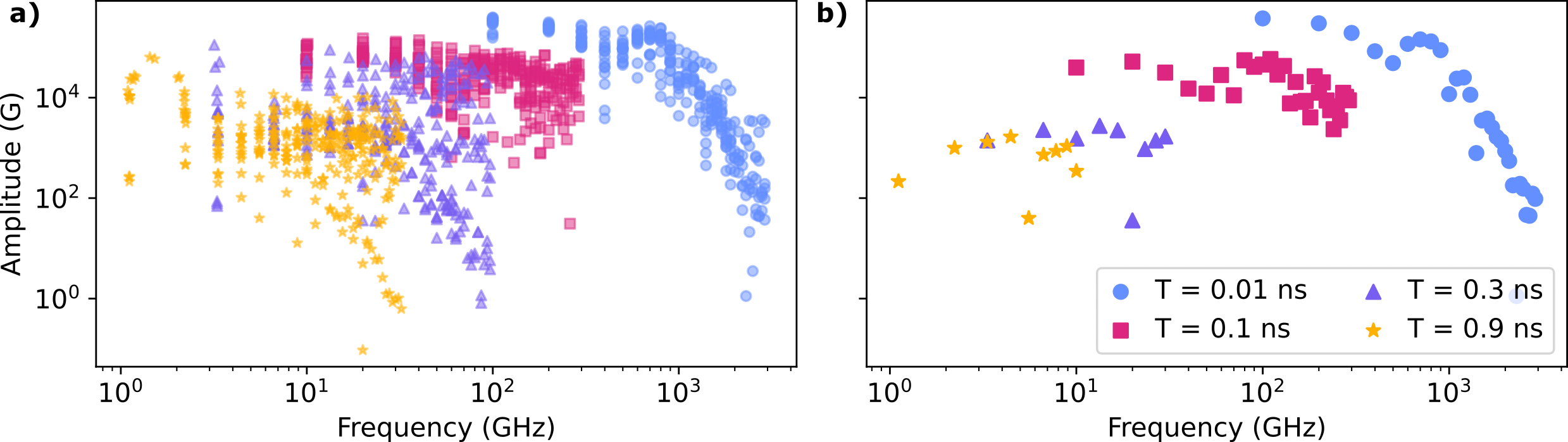}
    \caption{The optimal amplitude and frequency components of each sinusoidal basis function for all optimizations at various control duration. \textbf{a)} the optimal amplitude and frequency components of each optimal control pulse for a set of decreasing control duration. \textbf{b)} a subset of the data presented on the left, plotting only the frequency and amplitude components of the lowest infidelity control pulse found for each control duration.}
    \label{fig:amplitude_frequency}
\end{figure*}

Recently, experiments have been conducted to map the neighboring nuclear spin bath of an NV$^-$ center in great detail and extract the relevant intra- and inter-spin interaction strengths between nuclear and electronic spins \cite{abobeih_atomic-scale_2019}. It was found in Ref.~\cite{abobeih_atomic-scale_2019} that at least 27 $^{13}$C nuclei can be individually identified near the NV$^-$ center in a diamond sample with natural $^{13}$C abundance but it is anticipated that coupling to many more is possible \cite{onizhuk2023decoherence}. In our simulations we consider only two such nuclei but use the parameters found in Ref.~\cite{bradley_ten-qubit_2019,abobeih_atomic-scale_2019} to build an accurate model of an NV$^-$ center and its immediate environment. While a larger model would enable more accuracy, two nuclei were chosen to mitigate the memory requirements and enable a large number of quantum optimal control simulations to be run. 

We label basis states for this system in the eigenbasis of local spin-$z$ operators with a labeling $\ket{s_{z,e^-}, s_{z,^{14}N}, s_{z,^{13}C}, s_{z,^{13}C}}$ for the electron spin, nuclear spin, and the two carbon spins, respectively. For spin states with eigenvalue $+1$ and $-1$ we use $\ket{\uparrow}$ and $\ket{\downarrow}$. For example, for a spin-1 $S_z$ operator: $S_{z}\ket{\uparrow} = +1\ket{\uparrow}$, $S_{z}\ket{\downarrow} = -1\ket{\downarrow}$, $S_{z}\ket{0} = 0\ket{0}$.

The electron spin is used to define a qubit by applying a static external magnetic field in the $z$ direction, along the NV quantization axis. This lifts the degeneracy of the $\ket{m_s=\pm 1}$ states and allows one to define a magnetically addressable two-level subspace spanned by $\ket{m_s=0}$ and $\ket{m_s= -1}$. This is the standard qubit subspace used in most applications of NV$^-$ centers and the one we consider in this work. 

For our simulations, we assume that a static magnetic field of $B_z = 850$~G is constantly applied along the NV quantization axis defined by the zero-field splitting. This static bias ensures that the qubit transition frequency of about $0.49$~GHz is separated from the $\ket{m_s = 0 } \rightarrow \ket{m_s = +1}$ transition by about $5$~GHz. It has been shown in experiments that this enables precise addressing of the qubit subspace even during strong, fast driving \cite{fuchs_gigahertz_2009}. We assume $B_y(t) = 0$ and control the qubit by tuning the drive $B_x(t)$. We decompose this drive as the sum of parametrized sinusoidal functions and optimize the respective amplitude, frequency, and phase of each component, see Appendix~\ref{sec:methods} for additional details and methods.

Typical $\pi$-pulse controls for NV$^-$ center systems occur on the length of tens to hundreds of nanoseconds \cite{fuchs_gigahertz_2009,heremans_2016_control}. In this work we consider control duration less than $10$~ns. In this time regime, decoherence from both amplitude damping and spin-bath induced dephasing effects are negligible and we can consider only a unitary simulation without significant loss of accuracy \cite{rembold_introduction_2020}. 


\begin{figure*}
    \centering
    \includegraphics[width=\textwidth]{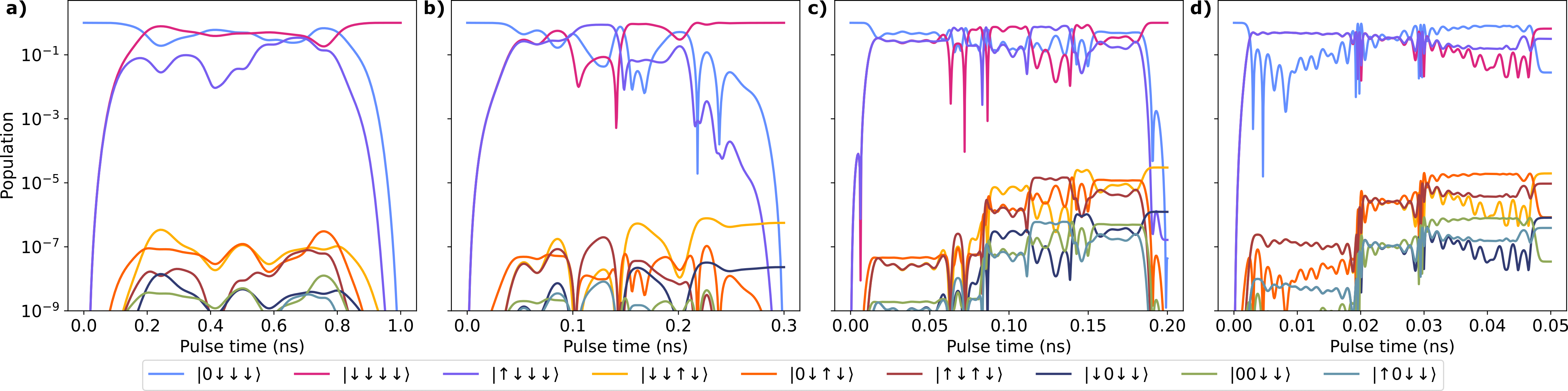}
    \caption{A set of plots visualizing the population dynamics generated by the most optimal control pulses. The initial state for all plots is $\ket{0\downarrow \downarrow \downarrow}$. A log scale is used on the y-axis in order to highlight the intermediate and final-time leakage populations. The final-time infidelities found via a high-accuracy simulation is approximately $3.07\times 10^{-10}$, $5.84\times 10^{-7}$, $3.14\times 10^{-5}$, and $0.35$ for control durations of $1.0$~ns, $0.3$~ns, and $0.2$~ns, and $0.05$~ns, respectively. The Hilbert space for the system under study has $36$ dimensions. However, for clarity, the only state populations plotted are those larger than $10^{-9}$.}
    \label{fig:population_dynamics_fast}
\end{figure*}

\section{Numerical results}\label{sec:numerical_results}

We implement an ensemble of quantum optimal control tasks to find population-inverting pulse waveforms for decreasing control duration. Our first set of numerical results is the infidelity of each pulse optimized for population inversion, shown in Fig.~\ref{fig:infidelity_amplitude}a for various number $N$ of pulse basis functions. We observe that there is an exponentially increasing infidelity with decreased control duration. Specifically, we note that control duration greater than 1~ns can achieve almost arbitrarily good control over the qubit subspace. 

However, at intermediate times, $10^{-1}\text{~ns} \leq t \leq 1$~ns, the optimal controls consistently obtain infidelities around $10^{-5}$. While not arbitrarily good control, this sub-nanosecond timescale is an interesting regime, because these fidelities are still relatively high from the standpoint of noisy intermediate-scale quantum systems and may be useful for several applications \cite{bradley_ten-qubit_2019}. 

We also vary the total number of basis functions, $N$, used to describe the pulse in the sub-ns regime. We observe that this only negligibly affects the attainable infidelity, suggesting that this infidelity limit is a fundamental property of the quantum system, not the assumed control decomposition.

In addition to the achievable fidelity, it is critical to examine the properties of the controls used to achieve these fidelities. In Fig.~\ref{fig:infidelity_amplitude}b we observe an exponentially increasing maximum pulse amplitude with decreasing time. While this is overall expected from quantum theory, the growth in pulse amplitude observed is more pronounced than what one would expect from purely analytical results. 

We plot for reference the required pulse amplitude for a $\pi$-pulse using a Gaussian envelope function and observe that the optimal pulses begin deviating significantly from the expected required amplitude around 1~ns. Within the two-level and rotating wave approximations, one expects from theory that the integrated area of the optimal pulse envelope will have integer multiples of $\pi$ area in order to induce population inversion. We observe this effect near $7-10$~ns where a hierarchy of optimal pulse amplitudes are found, separated roughly by a integer multiple. However, even our results for $10$~ns deviate from the ideal Gaussian $\pi$-pulse because we include all effects beyond the two-level and rotating wave approximations. 

The required pulse amplitude is only one limitation on the feasibility of a control pulse. In addition, the frequency components needed to generate the pulse must be quantified. This is visualized in Fig.~\ref{fig:amplitude_frequency} where all amplitude and frequency components (the $a_i,\omega_i$ in Eq.~\eqref{eq:ansatz}) for each optimization run are plotted on the left and only the most optimal pulse components are shown on the right. We observe that in addition to the exponential increase in amplitude, an exponential increase in frequency is required to represent the optimal pulses with decreasing control duration. 

Interestingly, the optimal controls for pulses within 1-10~ns lie within an experimentally achievable regime. For example, the required amplitudes for each frequency component are below 1000 Gauss and require frequencies below 10 GHz. This suggests that realizing true nanosecond control over NV$^-$ center system dynamics, while experimentally challenging, is not impossible.

We observe qualitatively a transition in infidelity in Fig.~\ref{fig:infidelity_amplitude}a below $1$~ns control times. To elucidate the mechanism of fidelity loss with decreasing control duration, we examine the population dynamics generated by the most optimal control pulses for pulse times below $1$~ns in Fig.~\ref{fig:population_dynamics_fast}.

First, we observe in Fig.~\ref{fig:population_dynamics_fast}a that for control approximately $0.5$~ns in length, the $\ket{m_s=+1}$ state is strongly populated at intermediate control duration. This indicates that the third level of the electron spin is actually being utilized coherently as a resource to generate the unitary evolution in the qubit subspace. Critically, at the final control duration all the population leaves the $\ket{m_s=+1}$ state, achieving the desired population inversion in the qubit subspace.

Next, we observe in Fig.~\ref{fig:population_dynamics_fast}b, with $T=0.3$~ns, that not all of the initial population returns to the qubit subspace at the final time. In fact, the two states $\ket{\downarrow \downarrow \uparrow \downarrow}$ and $\ket{\downarrow 0 \downarrow \downarrow}$ have final population $\mathcal{O}(10^{-6})$ and $\mathcal{O}(10^{-8})$, respectively. These two states represent partial flips of the nuclear spins of $^{13}C_1$ and $^{14}N$ induced by the control pulse.

Importantly, however, the final-time electron spin populations of $\ket{m_s = 0}$ and $\ket{m_s = +1} = \ket{\uparrow}$ are almost completely suppressed. This suggests that the loss in fidelity that occurs in Fig.~\ref{fig:infidelity_amplitude}a below $1$~ns is primarily due to population loss to the surrounding nuclear spin environment.

We now examine even shorter control duration. In Fig.~\ref{fig:population_dynamics_fast}c we observe that in addition to the final-time population of the nuclear spin environment, there is also comparable final-time population in the electron spin states, $\ket{m_s = 0}$ and $\ket{m_s = +1}$. However, both final-time populations are still below the population loss into the nuclear spin environment, which have increased slightly for the shorter duration pulse. 

In Fig.~\ref{fig:population_dynamics_fast}d, at a control duration of $0.05$~ns, we observe an evolution with infidelity of $0.35$. This infidelity is remarkably poor and indicates that the system is likely not controllable at such short times. Examining the population dynamics, we observe that nearly all of the initial population has become mixed into the other electron spin states $\ket{m_s = 0}$ and $\ket{m_s = +1}$ with only a relatively small increase in population lost to the nuclear spin bath. 

These observed population dynamics enable a simple explanation for the mechanism of short-time control and losses of fidelity seen in Fig.~\ref{fig:infidelity_amplitude}a. First, control pulses around $1$~ns coherently utilize the $\ket{m_s = +1}$ to induce the desired evolution in the qubit subspace. This is not surprising, as Fig.~\ref{fig:amplitude_frequency}(b) shows that the pulse has frequency components on the order of the splitting between the $\ket{m_s=-1}$ and $\ket{m_s=+1}$ transitions.  However, between approximately $0.3$~ns and $1.0$~ns control duration, population begins to transfer to the nuclear spin bath leading to fidelity loss. Below $0.3$~ns, the residual population of $\ket{m_s = 0}$ and $\ket{m_s = +1}$ induces further loss of fidelity until it becomes the primary source of fidelity loss below around $0.05$~ns.

\begin{figure*}
    \centering
    \includegraphics[width=\textwidth]{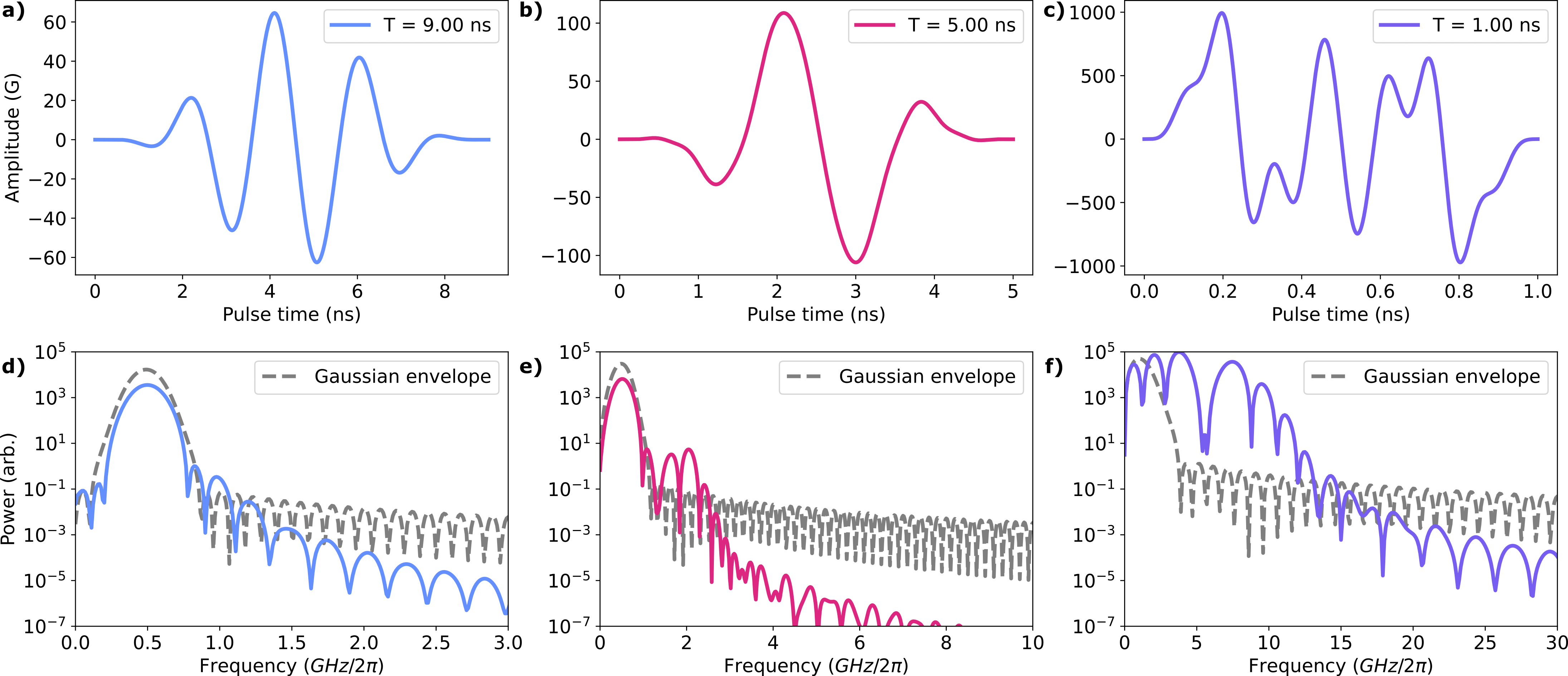}
    \caption{A set of the most optimal identified control pulses (\textbf{a,b,c}) and their respective power spectral densities (\textbf{d,e,f}) for various control duration. In the lower row, the power spectrum of a Gaussian envelope function with $\pi$ area for the same control duration is plotted for reference. The final-time infidelities found via a high-accuracy simulation is approximately $3.55\times 10^{-9}$, $4.28\times 10^{-12}$, and $5.07\times 10^{-11}$ for control duration $T=9.0$~ns, $T=5.0$~ns, and $T=1.0$~ns, respectively.}
    \label{fig:optimal_pulses}
\end{figure*}

Next, we examine the properties of the most experimentally relevant optimal pulses. Specifically, shown in Fig.~\ref{fig:optimal_pulses} are the time and frequency representations for the most optimal pulses in the laboratory frame of reference. We observe that the optimal solutions begin to slowly diverge from a Gaussian-like envelope under $10$~ns by adding more weight to higher frequency components. This happens slowly at first, but, for the optimal pulse at $1.0$~ns, the spectral density has broadened significantly. According to our calculations, population inversion occurring $<5$~ns would require 5-15~GHz of bandwidth to generate the control pulse. 

It is important to note the overall complexity of these optimal control pulses. Each is identified using a basis of only $10$ sinusoidal functions, totalling 30 parameters. In principle, this means each would require optimizing the same number of experimental parameters for calibration, suggesting that even with larger amplitudes, and larger bandwidths, the cost to calibrate these pulses should be similar \cite{werninghaus_leakage_2021}.

\section{Application in multipulse sequences}\label{sec:application}

NV$^-$ centers and the neighboring nuclear spin environment can be used for a variety of applications such as quantum computing and sensing. The effect of repeated $\pi$-pulses can be described in the language of multipulse sequences where ideal $\pi$-pulse sequences induce a filtering effect on the dynamical phase acquired by the qubit interacting with its environment \cite{biercuk_dynamical_2011}.

For quantum sensing, this filtering effect couples or decouples the qubit's phase evolution to specific frequencies in an external oscillating magnetic or electric field, thus providing a way to detect and characterize external signals \cite{biercuk_dynamical_2011,barry_sensitivity_2020}. The same mechanism is useful for quantum computing, where multipulse sequences can enhance entanglement rates between NV$^-$ center qubits and nuclear spins, allowing selective and robust quantum information processing \cite{abobeih_fault-tolerant_2022,bradley_ten-qubit_2019}. Additionally, multipulse sequences induce dynamical decoupling of the qubit from its environment, yielding longer coherence times in general \cite{dobrovitski_quantum_2013,anderson_five-second_2022}. 

In all of these applications the population inversion must be simultaneously high-fidelity, fast, and free from non-Markovian effects \cite{loretz_spurious_2015}. Non-Markovian dynamics arise from coherent exchange of quantum information between the system and its environment on timescales comparable to the dynamics of the system \cite{de_vega_dynamics_2017}. These effects can lead to complex, time-correlated errors in long multipulse sequences \cite{barry_sensitivity_2020}. Thus it is critical to quantify any non-Markovian effects that may arise from the optimal $\pi$-pulses found in this work.

\begin{figure*}
    \centering
    \includegraphics[width=\textwidth]{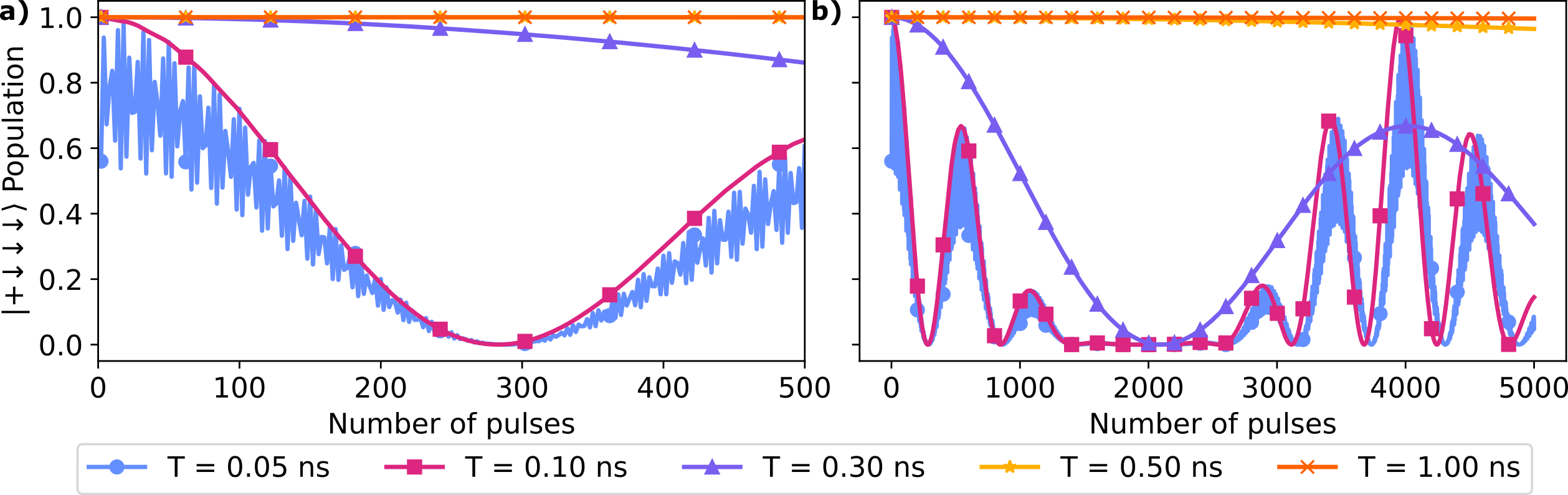}
    \caption{The population dynamics of $\ket{+ \downarrow \downarrow \downarrow}$ under a typical multipulse sequence of population inversions for various optimal controls. In an ideal case, the evolution induced in the qubit subspace during each control pulse should be unitary and involutory, thereby inducing no evolution after two $\pi$-pulses. Therefore, only the population at an even number of pulses is shown. Both plots are identical, but \textbf{a)} has an x-axis truncated at $500$ pulses to highlight the short-multipulse dynamics of the quantum state whereas \textbf{b)} shows the dynamics for a sequence up to $5000$ pulses. The legend labels the corresponding pulse duration.}
    \label{fig:dd_fidelity}
\end{figure*}

We probe the non-Markovianity of our optimal control pulses by simulating a multipulse sequence of repeated population inversion, i.e. repeatedly applying an $X$ gate to the qubit subspace. This is a realization of the Carr-Purcell
 sequence \cite{barry_sensitivity_2020,degen_quantum_2017}. Ideally, the evolution induced in the qubit subspace during each control pulse should be unitary and
involutory, thereby inducing no evolution after two $\pi$-pulses. In Fig.~\ref{fig:dd_fidelity} we show the population of the quantum state $\ket{+ \downarrow \downarrow \downarrow}$ as a function of number of pulses applied, focusing on sub-nanosecond optimal controls, where $\ket{+} = \frac{1}{\sqrt{2}} (\ket{m_s = 0} + \ket{m_s = -1})$ is a typical sensing state \cite{degen_quantum_2017}. 

Critically, in Fig.~\ref{fig:dd_fidelity}a we observe that, for pulses with duration above $0.3$~ns, population begins to decrease monotonically with increasing pulse number, suggesting that population leaving the qubit subspace does not return within about $500$ control pulses. However, when the number of pulses increases beyond this or when the pulse duration is shorter, we observe in Fig.~\ref{fig:dd_fidelity}b that there can be a coherent return of population into the qubit subspace, yielding non-Markovian effects. This population collapse and revival would dramatically erode the effectiveness of these pulses for applications such as dynamical decoupling or long-time multipulse sensing where it is common to use sequences of thousands of pulses \cite{barry_sensitivity_2020}.

We showed for the $T=0.05$~ns control duration in Fig.~\ref{fig:population_dynamics_fast}d that this pulse has significant residual coupling outside of the qubit subspace to the $\ket{m_s = +1}$ electron state. In a multipulse sequence, this will yield fast oscillations of population with number of pulses. However, pulses of slightly longer duration were shown in Fig.~\ref{fig:population_dynamics_fast} to primarily exchange population with the nuclear spins at a significantly lower rate. Therefore, in Fig.~\ref{fig:dd_fidelity}, we attribute the high frequency dynamics with population dynamics of the electron spin and low frequency population dynamics is attributed to the relatively slower population exchange with the nuclear spin environment.

\section{Conclusion}\label{sec:conclusion}

In this work we have explored the sub-nanosecond controllability of an NV$^-$ center electron spin qubit using high-accuracy numerical simulations and studied the feasibility of these controls for relevant applications. We have found an exponential increase in infidelity with decreasing control duration along with comparable exponential increases in the pulse amplitude and bandwidth required for control. Importantly, these amplitude and bandwidth requirements grow even faster than what we would expect analytically.

We identify a regime of arbitrarily good control with pulse times near and above $1$~ns in duration. In these control pulses, the third level of the NV$^-$ center electron spin $\ket{m_s = +1}$ is coherently used as a resource in order to generate fast, nearly unitary population inversion in the qubit subspace.

We also identify an intermediate regime of control duration yielding infidelity $\mathcal{O}(10^{-5})$ just below $1$~ns. We were able to determine that the loss of fidelity is due primarily to the coupling of the qubit to the neighboring nuclear spin environment. This begins to change with controls less than $0.2$~ns where the electron spin qubit cannot be controlled well and infidelities become unreasonably large.

Finally, we show that the optimal sub-nanosecond control pulses will actually yield non-Markovian effects when applied to multipulse sequences larger than about 500 pulses. This makes these short-time pulses potentially inadequate for tasks in quantum sensing and quantum information, even though their fidelity is relatively high. 

We conclude that it is possible to reduce the population inversion time to nearly 1~ns and still achieve arbitrarily good population inversion. If realized in the laboratory, these ultrashort controls could be used to substantially increase the accuracy, precision, and range of quantum sensing and information processing in NV$^-$ diamond systems. However, these controls seem to require maximum pulse amplitudes near $1000$~G and bandwidth near to $10$~GHz making their implementation challenging. Going further below this threshold reduces fidelity and increases control requirements exponentially.

The simulations performed in this work assumed a static magnetic field of $850$~G aligned with the NV$^-$ center principle axis. This is a large magnetic field and was used in previous work to create a well-isolated qubit system \cite{fuchs_gigahertz_2009}. However, our simulations have shown that coherent use of the third electron spin level $\ket{m_s = +1}$ is critical to fast control, indicating that a perfectly well-isolated qubit is not necessary and therefore large magnetic fields may not be required for ultrafast control. One straightforward route for continued work is to explore the impact of level splitting from the bias field, or even the impact of non-static bias fields on the controllability of NV$^-$ center qubits. This may be useful to further reduce the amplitude and bandwidth requirements for optimal pulses and enable near-term realizations in the laboratory. 

Looking forward, we found that below 1~ns the qubit begins to interact with the neighboring nuclear spins with the largest coupling due to the interactions with $^{13}C$ nuclei. Therefore,  material synthesis methods using isotopic purification of the $^{13}$C spin bath may provide a way to mitigate these deleterious couplings and achieve faster control. In future work, understanding the impact of material properties and synthesis strategies will be critical to push spin defect technologies towards their ultimate limits.

\section{Acknowledgements}

This material is based upon work supported by the U.S. Department of Energy Office of Science National Quantum Information Science Research Centers as part of the Q-NEXT center.

\appendix

\section{Methods}\label{sec:methods}

\subsection{Optimal control of NV center}\label{sec:qoc}

In this section we explain our definition of the control parametrization, constraints, and quantum optimal control task. First, we use the standard assumption that information processing is done in the reference frame rotating at the frequencies of the individual spins. 

We decompose the driving field as a sum of parametrized basis functions:

\begin{equation}
    b(t) = \sum_{i=1}^N a_i \sin(\omega_i t + \phi_i)
\end{equation}

where $a_i,\omega_i,\phi_i$ are the amplitude, frequency, and phase of the sinusoidal functions and $N$ determines the number of basis functions used. In this work we vary the total number of basis functions in order to understand optimal pulse complexity.

We require that the pulses have finite time, and therefore set boundary conditions that the pulse begins and ends at zero amplitude. This is enforced by modulating the function $b(t)$ with a flat-topped cosine function:

\begin{align}
    \Omega(t) = 
    \begin{cases} 
      \frac{1-\cos(\pi t/\tau_r)}{2}\Omega_m & 0\leq t \leq \tau_r \\
      \Omega_m & \tau_r \leq t \leq (\tau_c-\tau_r) \\
      \frac{1-\cos(\pi (\tau_c -t)/\tau_r)}{2} \Omega_m & (\tau_c - \tau_r) \leq t \leq \tau_c,
   \end{cases}
\end{align}
where $\tau_c = T$ is the total control duration, $\Omega_m=1$ scales the magnitude of the control pulse, and $\tau_r$ is the ramp time, which was chosen to be $0.3\tau_c$ to reduce spectral leakage \cite{tripathi_operation_2019}.

These two functions combine to yield our parametrization of the dynamical magnetic field:

\begin{equation}\label{eq:ansatz}
    B_x(\vec{\alpha},t) = \Omega(t)b(\vec{\alpha},t)
\end{equation}

where $\vec{\alpha} = (a_1,\omega_1,\phi_1,\dots,a_N,\omega_N,\phi_N)$ is the vector of parameters with total length $3N$. 

The global time evolution of the system for control duration $T$ and control parameters $\vec{\alpha}$ is formally written as

\begin{equation}
    U(\vec{\alpha},T) = \mathcal{T} \exp\bigg[ -\frac{i}{\hbar} \int_0^T d\tau H(\vec{\alpha},\tau) \bigg]
\end{equation}

While the global evolution is always unitary, the quantum dynamics within a subspace will generally not be. In order to identify controls that are unitary in the computational subspace, we measure the fidelity between the target population inversion in the qubit subspace,

\begin{align}
    X_{e} &= \ket{m_s=0}\bra{m_s=-1} \nonumber \\&+\ket{m_s=-1}\bra{m_s=0},
\end{align}

and the projection of the global final-time unitary evolution:

\begin{equation}
    u(\vec{\alpha},T) = P_q U(\vec{\alpha},T) P_q
\end{equation}

where $P_q$ is the projector onto the qubit subspace. 

We measure the infidelity between the two evolutions using the following function derived from the Hilbert-Schmidt norm:

\begin{equation}\label{eq:infid}
    g(\vec{\alpha},T) = 1- \frac{|\Tr(u(\vec{\alpha},T)X^\dagger_e)|^2}{2}.
\end{equation}

The infidelity function ranges from $0$ to $1$ and obtains a minimum when the subspace evolution $u(\vec{\alpha},T)$ is unitary and equivalent to $X_e$, up to a global phase. We identify optimal controls by minimizing the infidelity function.

Finally, it is important to note that the infidelity function is defined only at the final control duration $T$. When performing optimization this function will not penalize population leakage outside of the qubit subspace during the control duration so long as it returns to the qubit subspace at the conclusion of the pulse.

\subsection{Numerical methods}
We solve the optimal control problem using the Gradient Optimization of Analytic conTrols (GOAT) method \cite{machnes_gradient_2018}. We use the programming language Julia and various open-source packages \cite{Bezanson_Julia_A_fresh_2017}. A public release of the software can be found on a Github repository \cite{kairys_goat_2023}. Our implementation uses the Julia package DifferentialEquations.jl to solve the coupled GOAT equations of motion using a order 5/4 Runge-Kutta method with adaptive time stepping \cite{rackauckas2017differentialequations}. For the gradient-based control optimization of $\vec{\alpha}$, we use a limited-memory Broyden-Fletcher-Goldfarb-Shanno (L-BFGS) algorithm with a backtracking line-search method which are implemented in the Optim.jl package and LineSearches.jl package, respectively \cite{mogensen2018optim}. We limit each optimization to 1000 iterations of L-BFGS and define a stopping criteria when the infinity-norm of the gradient falls below 1e-9 or the relative change in the objective function is below 1e-8. For further details on the derivations of gradients via the GOAT algorithm we refer the reader to the original manuscript introducing GOAT, our previous work, and the package documentation \cite{machnes_gradient_2018,kairys_parametrized_2021, kairys_goat_2023}. 

All numerical data and associated codes are available upon request. 


\bibliographystyle{unsrt}
\bibliography{refs_static}

\end{document}